# Harmonic Generation in Multi-Resonant Plasma Films

M. A. Vincenti[1], D. de Ceglia[1], J. W. Haus[1,2], M. Scalora[3]

[1] *National Research Council - AMRDEC, Charles M. Bowden Research Laboratory, Redstone Arsenal - AL, 35898 – USA*

[2] *Electro-Optics Program, University of Dayton, 300 College Park, Dayton, OH, 45469, USA*

[3] *Charles M. Bowden Research Laboratory, AMRDEC, US Army RDECOM, Redstone Arsenal - AL, 35898, USA*

[*] *e-mail: maria.vincenti@us.army.mil*

**Abstract**

We investigate second and third harmonic generation in a slab of material that displays plasma resonances at the pump and its harmonic frequencies. Near-zero refractive indices and local field enhancement can deplete the pump for kW/cm$^2$ incident powers, without resorting to other resonant photonic mechanisms. We show that low-threshold, highly-efficient nonlinear processes are possible in the presence of losses and phase-mismatch in structures that are $10^4$ times shorter than typical KDP or LiNbO$_3$ crystals, for relatively low irradiance values.

The discovery of second harmonic generation (SHG) [1] was pivotal to the development of modern optics. During the last half century a plethora of devices and applications related to light generation have flourished, in large part brought about by the voluminous research on novel nonlinear materials [2] and the enormous potential of nonlinear processes [3-7]. The path toward efficient frequency conversion processes has thus retained its fundamental importance through the years and continues to be the focus of considerable basic research activity. High efficiency typically requires increasing the interaction length through phase-matching [3], quasi-phase-matching techniques [8], or engineering artificial structures that circumvent the lack of birefringence to compensate refractive index dispersion [4, 9]. For example, the unusually high density of states and field localization that characterizes periodic structures increases conversion efficiency [10, 11]. Less efficient, but equally important and useful for surface characterization and sensing are nonlinear processes originating from metallic surfaces. There, electric field enhancement triggered by the excitation of surface plasmons may suffice to detect harmonic processes arising from symmetry breaking at the surface, magnetic dipoles, electric quadrupoles, inner-core electrons, convective nonlinear sources and electron gas pressure [12, 13].

Recently, the focus has shifted to materials exhibiting near-zero effective permittivity [14]. While the preponderance of this work is still confined to the study of linear optical processes, these materials have also been predicted to exhibit enhanced harmonic generation [15-18] and optical instabilities [19, 20]. In this new setting nonlinear processes may be triggered by photonic [15, 16], plasmonic [18, 20], and plasma resonances [17, 18, 20]. In the latter scenario a singularity pushes the local field to achieve unusually large values due to the continuity of the displacement field component normal to the boundary [17, 21]. Nearly *all* natural materials display plasma resonances in the far infrared (LiF, $CaF_2$, $MgF_2$ or $SiO_2$), visible (Au, Ag, Cu), and ultraviolet (GaAs, GaP) frequency ranges [22] although in most cases absorption reduces field enhancement and abates the nonlinear response. However, glasses doped with absorbing dyes [17] and the introduction of gain material in artificial composites may significantly tame losses [23, 24].

In this paper we demonstrate that if plasma resonances are present *simultaneously* at the pump and its harmonic wavelengths*,* an unparalleled improvement of harmonic generation occurs compared to either plasma films having a single plasma resonance at the pump wavelength, or more traditional, phase-matched crystals like KDP or LiNbO$_3$. Near-zero permittivity and index values, along with an implicit phase-matching condition and pronounced field enhancement at the pump and its generated harmonic frequencies combine to increase the efficiency of harmonic generation, leading to pump depletion in sub-micron material lengths and irradiances of a few kW/cm$^2$, even in the presence of losses.

By way of example, we consider a homogenous slab of material of finite thickness $d$, surrounded by air and illuminated by a TM-polarized plane wave incident at an angle $\vartheta_i$ with respect to the $z$-axis, with electric field and wave-vector on the $x$-$z$ plane − inset of Fig. 1(a). The slab has relative permittivity $\varepsilon(\omega)$, which is modeled using classical Lorentz oscillators:

$$\varepsilon(\omega) = 1 - \sum_j \frac{\omega_{p_j}^2}{\omega^2 - \omega_{0j}^2 + i\omega\gamma_j} . \qquad (1)$$

The plasma frequencies are denoted by $\omega_{p_j}$, $\gamma_j$ are damping coefficients, $\omega_{0j}$ are resonance frequencies, $\omega$ is the angular frequency, and $i \equiv \sqrt{-1}$. The parameters are scaled with respect to the reference angular frequency $\omega_r = 2\pi c/\mu m$, where $c$ is the speed of light in vacuum. For simplicity we first assume a lossless scenario to establish ideal conditions, and then examine more realistic settings. The parameters $\omega_{p1} = 0.96835$, $\omega_{p2} = 0.56349$, $\omega_{01} = 0.25$, $\omega_{02} = 1.76$, and $\gamma_1 = \gamma_2 = 0$ produce two plasma resonances near the pump, $\lambda_\omega$ =1064nm, and second harmonic (SH), $\lambda_{2\omega}$ =532nm, wavelengths. Both wavelengths are slightly detuned from the plasma resonances so that the relative dielectric permittivities are: $\varepsilon_M = \varepsilon(\omega) = \varepsilon(2\omega) = 0.001$.

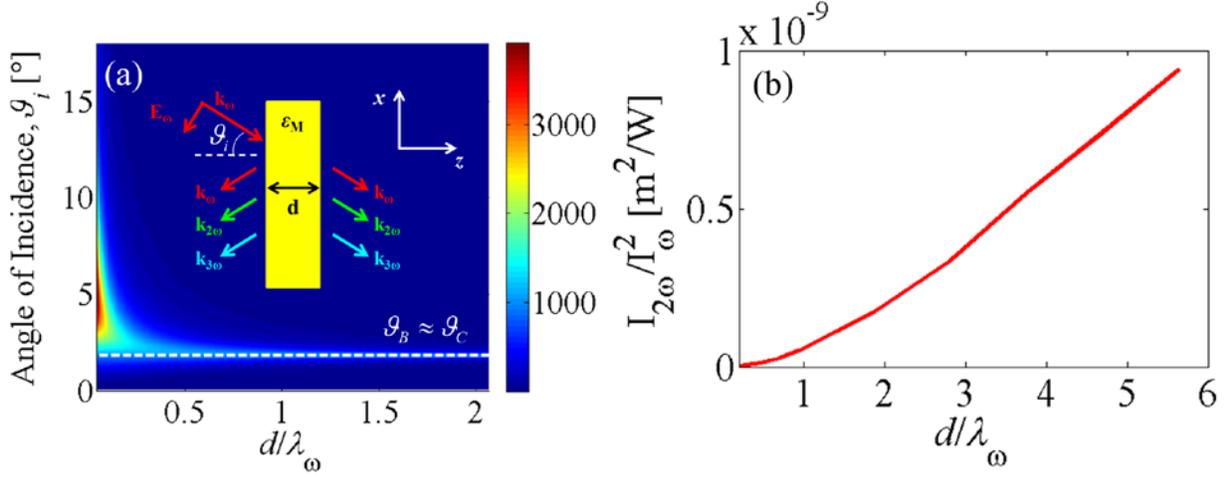

Fig. 1. (a) Field intensity map as a function of normalized slab thickness and incident angle; the enhancement approaches 8000 as $d \to 0$ (not shown); (inset) sketch of the system under investigation: a TM-polarized pump with electric field $\mathbf{E}_\omega$ and wave-vector $\mathbf{k}_\omega$ impinges on a slab of thickness $d$ at an angle $\vartheta_i$; (b) Total normalized conversion efficiency vs. slab thickness; we assume $\chi^{(2)} = 1 \, \text{pm/V}$ and $\vartheta_i = 1.8°$.

When a monochromatic plane wave impinges on an interface between a generic medium and a material with relative permittivity that *tends to zero* ($\text{Re}(\varepsilon(\omega)) \to 0^+$), the continuity requirement for the displacement field component normal to the surface causes the longitudinal (or normal) component of the electric field ($E_z$) inside the medium to become singular, i.e., $E_z = E_{z,0} \varepsilon_0 / \varepsilon_M$, where $E_{z,0}$ and $\varepsilon_0$ are the normal component of the electric field and permittivity, respectively, in the half-space of incidence. This occurs at Brewster or critical angle conditions [21]. In slabs of finite thickness $E_z$ may become singular by either: (i) reducing the thickness of the slab; or (ii) approaching normal incidence ($\vartheta_i \to 0$) [21]. If, on the other hand, $\text{Re}(\varepsilon(\omega))$ is fixed to a *near-zero* value, the electric field is *not singular* but may be significantly enhanced by exploiting the tunneling of evanescent waves and multiple reflections that occur for thinner slabs. Electric field intensity enhancement, calculated as $\max(|E_{z\phi}|^2 / |\mathbf{E}_\omega|^2)$, is shown in Fig. 1(a) as a function of normalized slab thickness $d/\lambda_\omega$ and incident angle $\vartheta_i$. In the figure

enhancement is largest for incident angles greater than the critical angle $\vartheta_C = \sin^{-1}\sqrt{\varepsilon_M}$ [21] – white dashed curve labeled $\vartheta_C$ in Fig. 1(a). If $\vartheta_i > \vartheta_C$ evanescent waves tunnel through the slab and, together with multiple reflections, further enhance $E_z$. Tunneling is more pronounced for thinner slabs and the enhancement eventually approaches a value of 8000 as $d \to 0$. Fig. 1(a) shows that for *near-zero* permittivity the Brewster angle, $\vartheta_B = \tan^{-1}\sqrt{\varepsilon_M}$, is approximately equal to the critical angle $\vartheta_C$. Since $\varepsilon(\omega) = \varepsilon(2\omega)$, similar field enhancement is expected for input waves tuned at $\lambda_\omega$ and $\lambda_{2\omega}$.

When the pump and its harmonic(s) are enhanced, nonlinear processes become favored at low irradiances even in sub-wavelength structures. Second and third harmonic (TH) generation are accounted for by writing a nonlinear polarization in the time domain as [25]:

$$P_k^{\rm NL} = \varepsilon_0 \sum_{l,m} \chi_{klm}^{(2)} E_l E_m + \varepsilon_0 \sum_{l,m,n} \chi_{klmn}^{(3)} E_l E_m E_n, \qquad (2)$$

where $\varepsilon_0$ is the vacuum electric permittivity, $k,l,m,n$ are the Cartesian coordinates, $P_k^{\rm NL}$ is the total nonlinear polarization in the *k*-th Cartesian direction. We assume a near-instantaneous nonlinear response and that only the diagonal components of the tensors $\chi_{klm}^{(2)}$ and $\chi_{klmn}^{(3)}$ are non-zero, such that $\chi_{xxx}^{(2)} = \chi_{yyy}^{(2)} = \chi_{zzz}^{(2)} = \chi^{(2)} = 1$ pm/V and $\chi_{xxxx}^{(3)} = \chi_{yyyy}^{(3)} = \chi_{zzzz}^{(3)} = \chi^{(3)} = 10^{-20}$ m$^2$/V$^2$. For a TM-polarized field that is a superposition of the pump, second and third harmonics, the nonlinear polarizations may be expressed in terms of space and time dependent, complex envelope functions [13]:

$$\begin{aligned} P_{\omega,k}^{\rm NL} = &\, 2\varepsilon_0 \chi^{(2)} \left( E_{2\omega,k}^* E_{3\omega,k} + E_{\omega,k}^* E_{2\omega,k} \right) \\ &+ 3\varepsilon_0 \chi^{(3)} \left\{ \left( |E_{2\omega,k}|^2 + 2|E_{3\omega,k}|^2 + 2|E_{\omega,k}|^2 \right) E_{\omega,k} + E_{2\omega,k}^2 E_{3\omega,k}^* + E_{\omega,k}^* E_{\omega,k}^{*2} \right\}, \end{aligned} \qquad (3)$$

$$P_{2\omega,k}^{NL} = \varepsilon_0 \chi^{(2)} \left( E_{\omega,k}^2 + 2E_{\omega,k}^* E_{3\omega,k} \right)$$
$$+ 3\varepsilon_0 \chi^{(3)} \left\{ \left( |E_{3\omega,k}|^2 + 2|E_{\omega,k}|^2 + 2|E_{2\omega,k}|^2 \right) E_{2\omega,k} + 2E_{3\omega,k}^* E_{\omega,k} E_{\omega,k} \right\}, \quad (4)$$

$$P_{3\omega,k}^{NL} = 2\varepsilon_0 \chi^{(2)} E_{2\omega,k} E_{\omega,k}$$
$$+ \varepsilon_0 \chi^{(3)} \left\{ E_{2\omega,k}^3 + 3E_{\omega,k}^2 E_{\omega,k}^* + \left( 6|E_{\omega,k}|_{3\omega}^2 + 3|E_{\omega,k}|_{\omega}^2 + 6|E_{3\omega,k}|^2 \right) E_{\omega,k} \right\}. \quad (5)$$

At low pump intensities self- and cross-phase modulation terms, as well as down-conversion terms do not contribute. In this regime, in Fig. 1(b) we plot the total normalized SH conversion efficiency $I_{2\omega}/I_\omega^2$ from a slab of variable thickness and incident angle $\vartheta_i \approx \vartheta_B \approx \vartheta_C = 1.8°$ when pump and SH fields are tuned near plasma resonances. The pump field intensity inside the slab is enhanced by a factor of 1000 regardless of its thickness (white dashed line in Fig. 1(a)). Choosing $\vartheta_i = 1.8°$ does not maximize field enhancement, but the simultaneous boost imparted to incident and generated fields, implicit phase-matching, and correspondingly near-zero indices of refraction induce pump depletion for relatively small peak powers. For example, SH conversion efficiency of a plasma slab 1 μm thick, illuminated with $I_\omega = 20$ kW/cm$^2$ at $\vartheta_i = 1.8°$ is 1%. As another example, we find that a lossless, doubly-resonant plasma film 6 μm thick pumped with $I_\omega = 100$ kW/cm$^2$ yields a 30% SH conversion efficiency. The same efficiency is achieved in a phase-matched KDP crystal 45 cm thick, under similar pumping conditions.

A multi-resonant plasma film is implicitly phase-matched. However, phase-matching is important only for relatively thick films. Under exact phase-matching conditions the coherence length is infinite ($L_c \equiv 2/\Delta k$, with $\Delta k = k_{2\omega} - 2k_\omega$ and $k_{\omega,2\omega} = 2\pi \operatorname{Re}(n_{\omega,2\omega})/\lambda_{\omega,2\omega}$, where $n_{\omega,2\omega} = \sqrt{\varepsilon_{\omega,2\omega}}$). *For sub-wavelength films, the major contribution to the nonlinear process* *comes from the condition* $n_\omega \approx n_{2\omega} \approx 0$. This result is inferred from a fresh re-evaluation of the normalized conversion efficiency [25]:

$$\frac{I_{2\omega}}{I_\omega^2} = \frac{2(\chi^{(2)}\omega d)^2}{n_\omega^2 n_{2\omega} \varepsilon_0 c^2} \left(\frac{\sin(\Delta k d/2)}{\Delta k d/2}\right)^2. \qquad (7)$$

Eq.(7) reveals that conversion efficiency becomes singular when $n_\omega$ and/or $n_{2\omega}$ approach zero. In Fig. 2(a) we plot the normalized conversion efficiency vs. $n_\omega$ and $n_{2\omega}$ obtained using Eq. (7) for a 2 μm long bulk medium. The map shows that conversion efficiency is largest when $n_\omega \to 0$ and/or $n_{2\omega} \to 0$. While these conditions dominate over the influence of phase matching, they cannot overcome the effect of wave vector mismatch, where efficiency vanishes. However, mere inspection of Eq.(7) does not suffice to describe a complete, realistic scenario, because it falls short in at least two respects: (i) it is valid only for bulk media pumped at normal incidence ($\vartheta_i=0$); (ii) it is invalidated if the pump becomes depleted. Under realistic conditions optimal coupling of wave incident obliquely is achieved only for a specific angle of incidence $\vartheta_i$ and finite thickness $d$ (see Fig. 1(a)). Finally, when $n_\omega \to 0$ and/or $n_{2\omega} \to 0$ pump depletion can occur for relatively small incident fields. Therefore, accurate estimates of conversion efficiency should always be done with an analysis that takes Eqs.(3-5) into account. For these reasons we used three different methods to calculate conversion efficiencies in slabs of finite thickness: (i) continuous wave (CW) analysis via a Finite Element Method (Comsol Multiphysics), and pulsed dynamics using: (ii) a fast Fourier transform beam propagation method, and (iii) finite difference time domain method.

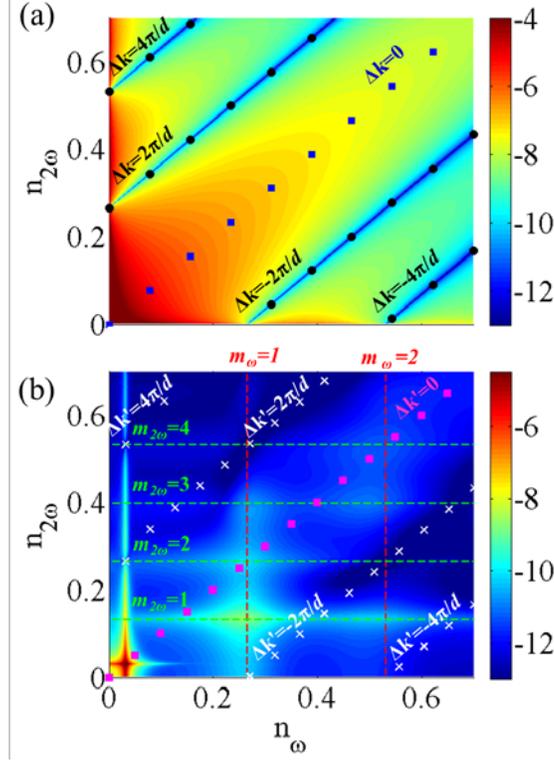

Fig. 2. (a) Normalized SH conversion efficiency ($I_{2\omega}/I_\omega^2$) on a logarithmic color scale for a bulk medium 2μm long calculated using Eq. (7), assuming normal incidence and $\chi^{(2)} = 1\,\text{pm/V}$; Blue, square markers indicate the phase-matching condition; Black, circle markers indicate wave vector mismatch conditions. (b) Total (forward plus backward) normalized SH conversion efficiency ($I_{2\omega}/I_\omega^2$) on a logarithmic color scale for a slab $d=2$μm thick, calculated by means of full-wave simulations (CW analysis) assuming $\chi^{(2)} = 1\,\text{pm/V}$ and $\vartheta_i = 1.8°$; Purple, square markers indicate the phase-matching condition; White, cross markers indicate wave vector mismatch conditions; red and green dashed lines indicate Fabry-Perot resonances for $\lambda_\omega$ and $\lambda_{2\omega}$, respectively.

As was done for the bulk medium, phase matching and wave vectors mismatch conditions are highlighted in the new efficiency map (Fig. 2(b)). The phase matching/mismatch conditions Δk' in Fig. 2(b) do not coincide with those in Fig. 2(a) because we are incident at an angle $\vartheta_i$=1.8°. Although it is still possible to identify the beneficial effects of phase matching and the detrimental effect of wave vector mismatch, one may easily ascertain that enhancement of SHG occurs when either singly- or doubly-resonant conditions are met. Fabry-Perot resonances for $\lambda_\omega$ or $\lambda_{2\omega}$ are excited when $d = (m_{\omega,2\omega} \lambda_{\omega,2\omega})/(2n_{\omega,2\omega})$, with $m_{\omega,2\omega} = 1,2,3...$ for the pump and the SH. Even though SH conversion efficiency is largest near a double *plasma* resonance, efficiency is generally maximized when Brewster/critical incidence condition is satisfied for the incident field − bright red region in Fig. 2(b), i.e, when the condition $\vartheta_i \approx \vartheta_B \approx \vartheta_C$ is met. Similarly large conversion efficiencies occur in a bulk medium only when $n_\omega \to 0$ and/or $n_{2\omega} \to 0$ − Fig. 2(a).

A doubly-resonant plasma film continues to show remarkable conversion efficiency even in the presence of losses and phase-mismatch. Using a slightly modified set of parameters to introduce losses ($\omega_{p1} = 0.957249$, $\omega_{p2} = 0.517856$, $\omega_{01} = 0.25$, $\omega_{02} = 1.78$, $\gamma_1 = \gamma_2 = 0.0001$) we obtain $\varepsilon_\omega = 0.001 + i0.00013$ and $\varepsilon_{2\omega} = 0.001 + i0.00039$. The main result is that the new dispersion profile and relative permittivities modify the electric field intensity enhancement as a function of slab thickness and angle of incidence, so that conversion efficiencies obtained without losses may be recovered at slightly larger angles of incidence and thinner slabs. For example, if we introduce losses, keep $\vartheta_i = 1.8°$, and choose slab thickness $d$ = 200 nm, the field intensity enhancement at $\lambda_\omega$ drops from ~1000 without losses to ~800 with losses. SH conversion efficiency is impacted by decreased local fields, and reduced from $10^{-8}$ of the lossless case to $10^{-9}$ in the lossy scenario for an irradiance of 1 W/cm$^2$. Conversion efficiency may be completely recovered by reducing the thickness to $d$ = 50 nm and by increasing the incident angle to $\vartheta_i = 4.4°$ with 1 W/cm$^2$ pump irradiance. This new combination of small thickness and larger incident angle restores a field intensity enhancement of ~1000 for the pump, hence the compensation in SH efficiency. As shown in Fig. 1(a), thinner slabs facilitate evanescent wave tunneling, and increase the optimal field enhancement angle. Losses also broaden the angular region where field intensity is boosted, regardless of slab thickness. These effects foster an electric field enhancement that compensates losses, leading to the same efficiencies achieved in the lossless scenario. Fig. 3(a) shows SH conversion efficiency as a function of input irradiance with and without losses for the same thickness and angle of approach. The figure shows the transition between undepleted and depleted pump regimes, saturation, and the influence of pulse width. The pulsed efficiency curves approach the CW results as pulse duration, τ, increases. At large intensities self- and cross-phase modulation effects become dominant and alter the permittivity of the slab. As a result, the slab becomes *detuned* from optimal coupling conditions ($\vartheta_i \approx \vartheta_B \approx \vartheta_C$), and conversion efficiency decreases regardless of losses or pulse width.

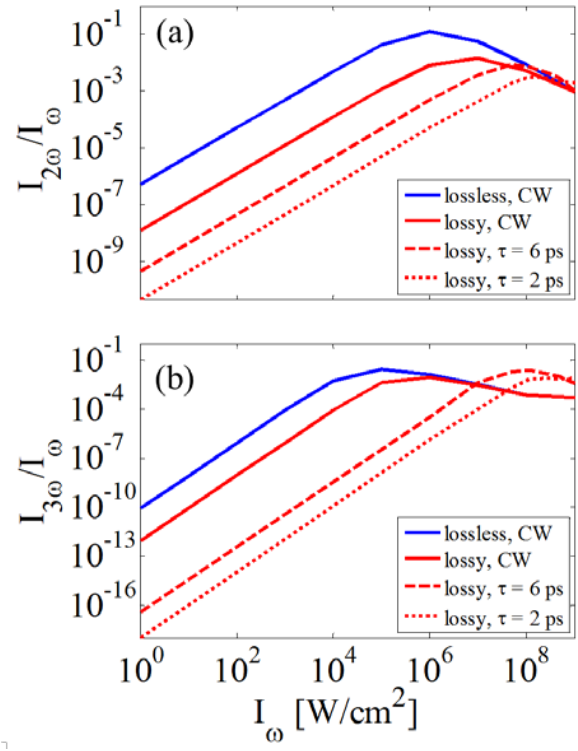

Fig. 3. (a) SH conversion efficiency vs. input irradiance for a 1μm-thick film and incident angle $\vartheta_i = 1.8°$ for: lossless CW (blue, solid curve), lossy CW (red, solid curve), τ = 2ps (red, dotted curve) and τ = 6ps (red, dashed curve) pulses for lossy film; (b) Same as (a) for THG.

Similar considerations are applicable to third harmonic generation (THG) if one considers a slab of material with plasma resonances close to the fundamental and TH frequencies. The amplitude of the TH field becomes comparable to the pump field, so that none of the self-phase modulation or down-conversion terms in Eqs. (3-5) can be neglected. In this lossless, doubly-resonant scenario TH conversion efficiency for $\vartheta_i = 1.8°$ and slab thickness $d = 1$ μm is ~3% for 100 kW/cm$^2$ input irradiance (blue, solid curve in Fig. 3(b)). We can estimate the impact of losses on THG by introducing slightly modified parameters ($\omega_{p1} = 0.956693$, $\omega_{p2} = 0.851393$, $\omega_{01} = 0.25$, $\omega_{02} = 2.67$, $\gamma_1 = \gamma_2 = 0.0001$) that yield $\varepsilon_\omega = 0.001 + i0.00013$ and $\varepsilon_{3\omega} = 0.001 + i0.00031$. Under these circumstances TH conversion efficiency is reduced as shown in Fig. 3(b) (red, solid curve). For irradiances larger than 1 MW/cm$^2$ self- and cross-phase modulation once again dominate the dynamics worsening tuning conditions. Similarly to what occurs for SHG, conversion efficiency flattens regardless of losses or pulse width.

In conclusion we have studied SHG and THG in slabs of materials that display multi-plasma resonances. Nonlinear processes are driven by the simultaneous enhancement of pump and harmonic electric fields in the absence of photonic resonances, and are characterized by pump depletion at relatively small irradiances even in the presence of losses and phase-mismatch. It is therefore evident that a novel path to efficient harmonic generation is not only realistic, but may well surpass the performance of commonly used nonlinear crystals, such as KDP or LiNbO$_3$.

**Acknowledgement**

This research was performed while the authors M. A. Vincenti, D. de Ceglia and J. W. Haus held a National Research Council Research Associateship award at the U.S. Army Aviation and Missile Research Development and Engineering Center.
**References**

1. P. A. Franken, A. E. Hill, C. W. Peters, and G. Weinreich, "Generation of Optical Harmonics," Physical Review Letters **7**, 118-119 (1961).
2. J. Zyss, *Molecular Nonlinear Optics: Materials, Physics, and Devices* (Academic Press, 1994).
3. J. A. Armstrong, N. Bloembergen, J. Ducuing, and P. S. Pershan, "Interactions between Light Waves in a Nonlinear Dielectric," Physical Review **127**, 1918-1939 (1962).
4. N. Bloembergen and A. J. Sievers, "Nonlinear optical properties of periodic laminar structures," Applied Physics Letters **17**, 483-486 (1970).
5. M. M. Fejer, "Nonlinear Optical Frequency Conversion," Physics Today **47**, 25-32 (1994).
6. E. C. Cheung, K. Koch, and G. T. Moore, "Frequency upconversion by phase-matched sum-frequency generation in an optical parametric oscillator," Opt. Lett. **19**, 1967-1969 (1994).
7. B. J. Eggleton, R. E. Slusher, C. M. de Sterke, P. A. Krug, and J. E. Sipe, "Bragg Grating Solitons," Physical Review Letters **76**, 1627-1630 (1996).
8. M. M. Fejer, G. A. Magel, D. H. Jundt, and R. L. Byer, "Quasi-phase-matched second harmonic generation: tuning and tolerances," Quantum Electronics, IEEE Journal of **28**, 2631-2654 (1992).
9. J. P. van der Ziel and M. Ilegems, "Second harmonic generation in a thin AlAs-GaAs multilayer structure with wave propagation in the plane of the layers," Applied Physics Letters **29**, 200-202 (1976).
10. J. Trull, R. Vilaseca, J. Martorell, and R. Corbalán, "Second-harmonic generation in local modes of a truncated periodic structure," Opt. Lett. **20**, 1746-1748 (1995).
11. M. Scalora, M. J. Bloemer, A. S. Manka, J. P. Dowling, C. M. Bowden, R. Viswanathan, and J. W. Haus, "Pulsed second-harmonic generation in nonlinear, one-dimensional, periodic structures," Physical Review A **56**, 3166-3174 (1997).
12. M. Scalora, M. A. Vincenti, D. de Ceglia, V. Roppo, M. Centini, N. Akozbek, and M. J. Bloemer, "Second- and third-harmonic generation in metal-based structures," Physical Review A **82**, 043828 (2010).
13. M. Vincenti, D. de Ceglia, V. Roppo, and M. Scalora, "Harmonic generation in metallic, GaAs-filled nanocavities in the enhanced transmission regime at visible and UV wavelengths," Optics Express **19**, 2064-2078 (2011).
14. M. Silveirinha and N. Engheta, "Tunneling of Electromagnetic Energy through Subwavelength Channels and Bends using ε-Near-Zero Materials," Physical Review Letters **97**, 157403 (2006).
15. A. Ciattoni, C. Rizza, and E. Palange, "Extreme nonlinear electrodynamics in metamaterials with very small linear dielectric permittivity," Physical Review A **81**, 043839 (2010).
16. A. Ciattoni and E. Spinozzi, "Efficient second-harmonic generation in micrometer-thick slabs with indefinite permittivity," Physical Review A **85**, 043806 (2012).
17. M. A. Vincenti, D. de Ceglia, A. Ciattoni, and M. Scalora, "Singularity-driven second- and third-harmonic generation at ε-near-zero crossing points," Physical Review A **84**, 063826 (2011).
12